\documentclass[manuscript]{revtex4}
\usepackage{graphicx}
\begin{document}

\title{On the interaction of a single-photon wave packet with an excited atom}
\author{P.V. Elyutin}
\email{pvelyutin@mtu-net.ru} \affiliation {Department of Physics,
Moscow State University, Moscow 119991, Russia}
\date{\today}

\begin{abstract}
The interaction of a single-photon wave packet with an initially excited two-level atom in free space is studied in semiclassical and quantum approaches.  It is shown that the final state of the field does not contain doubly occupied modes.  The process of the atom's transition to the ground state may be accelerated, decelerated or even reversed by the incoming photon, depending on parameters.  The spectrum of emitted radiation is close to the sum of the spectrum of the incoming single-photon wave packet and the natural line shape, with small and complicated deviations.

\vspace{10mm} PACS numbers: {42.50.Ct, 32.70.Jz, 42.55.Ah }
\end{abstract}
\maketitle

\section{Introduction}

The main purpose of this paper is a calculation of the kinetics of population of atomic levels and of the spectrum of light emitted as the result of the interaction of a single-photon wave packet with an excited atom.  The problem concerns an elementary process of interaction of light with matter that plays an important role in the laser theory.  The experiments with one-atom, one-photon setup can be traced back as far as 1985  \cite{MWM85}; with the techniques of today it seems possible to study the process in the experiment directly.

In Sec. 2 the simple classical model of the packet is introduced and used along with the quantum model of the atom to obtain the estimate of the difference of population induced by the photon.  This difference comes to be very small (in typical situations) and can change its sign.  In Sec. 3 the equations of a fully quantum model are derived.  They are solved in the approximation that starts with the assumption of the negligibility of the photon's influence on the atomic evolution.  The solutions, that are obtained in the form of quadratures, describe the kinetics of probabilities of the atomic states and the spectrum of the emitted radiation.  In Sec. 4 these data are visualized with the detailed study of a one-dimensional numerical example.  The discussion of discrepancies between the semiclassical solution and the simplest approximation of the quantum approach is also given in this section.  The established properties of the interaction of a single-photon wave packet with an initially excited two-level atom in free space are discussed in Sec. 5 in the context of the ongoing discussions of the nature of the stimulated emission.

\section{ Semiclassical approach }

The classical model that represents a photon (more precisely, a single-photon wave packet) as a limited in space pulse of the quasimonochromatic classical radiation is well known \cite{W67, BK94}.  The studies of the spatial distribution of the energy of the electromagnetic field of a photon in the framework of the quantum theory have been carried out actively in the recent years \cite{BB94, S95, DHB00}.

From the energy density considerations one can equate the total energy of the field of the pulse to the energy of the photon $\hbar \omega $, where $\hbar $ is the Planck's constant and $ \omega $ is the photon frequency: 
\begin{equation}\label{1}
\frac{1}{{8\pi }}\int {\left[ {{\bf{E}}\left( {{\bf{r}},t} \right)^2  + {\bf{H}}\left( {{\bf{r}},t} \right)^2 } \right]} d{\bf{r}} = \hbar \omega .
\end{equation}
Here ${\bf{E}}\left( {{\bf{r}},t} \right)$  and ${\bf{H}}\left( {{\bf{r}},t} \right)$ are the vectors of the electric and magnetic fields respectively.

In this section we shall threat the linearly polarized pulse of the Gaussian shape,  propagating in the positive direction of the $OX$  axis:
\begin{equation}\label{2}
{\bf{E}}\left( {{\bf{r}},t} \right) = {\bf{e}}{\mathcal{E}}_0 \exp \left[ { - \frac{{\left( {x - ct} \right)^2 }}{{4l^2 }}} \right]\cos \left[ {k\left( {x - \Lambda } \right) - \omega t} \right],
\end{equation}
where $\bf{e}$ is the polarization vector, ${\mathcal{E}}_0$ is the pulse amplitude (maximal value of the field), $l$ is the length of the pulse, $\Lambda$ is the initial displacement of the packet's center and $c$ is the speed of light.  We shall assume that the field is restricted in the transverse directions to some domain with a section area $S$; the exact form of the transverse structure will not be needed.

From Eq. (\ref{1}) we have
\begin{equation}\label{3}
{\mathcal{E}}_0  = \left( {8\pi } \right)^{{1 \mathord{\left/
 {\vphantom {1 4}} \right.
 \kern-\nulldelimiterspace} 4}} \sqrt {\frac{{\hbar \omega }}{{lS}}}.
\end{equation}

We shall study the action of this packet on a two-level atom, located at the origin of the coordinate system.  The atom will be represented by a two-level system with the transition frequency $\omega _0 $ and the dipole transition matrix element $d$ along the field direction.  The evolution of this system under the influence of the harmonic external field with the carrier frequency $\omega$ and the envelope ${\mathcal{E}}\left( t \right)$ (${\mathcal{E}}_0  = \max {\mathcal {E}}\left( t \right)$) can be described by the equations for the components of the Bloch vector  (in the rotating wave approximation (RWA) and in the rotating frame of reference), that have in general case the following form \cite{AE75}:
\begin{equation}\label{4}
\dot u + \Gamma _2 u =  - \Delta v,
\end{equation}
\begin{equation}\label{5}
\dot v + \Gamma _2 v = \Delta u + \Omega w,
\end{equation}
\begin{equation}\label{6}
\dot w + \Gamma _1 \left( {w - w_0 } \right) =  - \Omega v.
\end{equation}
Here $u$ and $v$ are the transverse components and the population difference $w$ is the longitudinal component of the Bloch vector, $w_0$ is the population difference at equilibrium, $\Gamma _2$ and $\Gamma _1$ are the rates of the transverse and longitudinal relaxations correspondingly, $\Delta  = \omega  - \omega _0 $  is the frequency detuning, and the Rabi frequency
\begin{equation}\label{7}
\Omega \left( t \right) = \frac{{d{\mathcal{E}}\left( t \right)}}{\hbar }.
\end{equation}
We shall limit ourselves to the resonant case $\Delta  = 0$.  In this case the system reduces to two last equations, that for our problem should be written as
\begin{equation}\label{8}
\dot v + \gamma v = \Omega \left( t \right)w,
\end{equation}
\begin{equation}\label{9}
\dot w + 2\gamma \left( {w + 1} \right) =  - \Omega \left( t \right)v.
\end{equation}

We have taken into account that for the spontaneous emission $2\Gamma _2  = \Gamma _1  = 2\gamma $, and the equilibrium corresponds to the system being in a ground state (no thermal noise).  The time-dependent Rabi frequency follows the pulse envelope Eq. (\ref{2}), and its form can be written as a Gaussian,
\begin{equation}\label{10}
\Omega \left( t \right) = \Omega _0 \exp \left[ { - \frac{{\left( {t - T} \right)^2 }}{{4\tau ^2 }}} \right],
\end{equation}
where $\Omega _0  = \max \Omega \left( t \right)$, $\tau  = {l \mathord{\left/ {\vphantom {l c}} \right. \kern-\nulldelimiterspace} c}$ is the pulse duration and $T = {{ - \Lambda } \mathord {\left/ {\vphantom {{ - \Lambda } c}} \right. \kern-\nulldelimiterspace} c}$ is the arrival time of the pulse peak to the point of location of the atom.  The spectrum of this pulse has the width $\delta  = \tau ^{ - 1} $.  The initial conditions for Eqs. (8,9) are: $v\left( 0 \right) = 0$, $w\left( 0 \right) = 1$, that is, at the initial moment the atom is in the excited state.

To find the proper approximation we turn to the numerical estimates for a typical situation.  For the atomic parameters we choose the transition frequency $\omega _0  = 3.54 \cdot 10^{15} \,{\rm{s}}^{ - 1} $ and the atomic unit of the dipole moment $d = ea_0  = 2.42 \cdot 10^{ - 18} \,{\rm{CGS}}$.  For these parameters the rate of the spontaneous emission is 
\begin{equation}\label{11}
\Gamma _1  = 2\gamma  = \frac{{4d^2 \omega _0^3 }}{{3\hbar c^3 }} = 1.34 \cdot 10^7 \,{\rm{s}}^{ - 1}.
\end{equation}
We take the duration of the pulse $\tau  = 1\,\,{\rm{ns}}$ and the area of the transverse section of the pulse $S = 5 \cdot 10^{ - 3} \,\,{\rm{cm}}^2 $.  Then for the maximal amplitude of the field we have ${\mathcal{E}}_0  = 1.18 \cdot 10^{ - 5} \,\,{\rm{Gs}}$, and for the maximal Rabi frequency $\Omega _0  = {{d{\mathcal{E}}_0 } \mathord{\left/ {\vphantom {{dE_0 } \hbar }} \right. \kern-\nulldelimiterspace} \hbar } = 2.56 \cdot 10^4 \,\,{\rm{s}}^{ - 1}  = 1.91 \cdot 10^{ - 3} \Gamma _1 $.  The Rabi frequency is small in comparison with the relaxation rate, and the system is overdamped.

Since the influence of the pulse on the atom is small (${\Omega _0  \ll \gamma }$), does not last long (${\gamma \tau  \ll 1}$) and commences around the arrival moment $T$, we replace in the RHS of Eq. (\ref{8}) the function $w\left( t \right)$ by its value at the arrival time $w\left( T \right)$.  Then for the change of the population difference under the influence of the pulse we have 
\begin{equation}\label{12}
\Delta w =  - \frac{{w\left( T \right)}}{2}\left( {\int\limits_0^\infty  {\Omega \left( t \right)dt} } \right)^2 .
\end{equation}
The absence of $\gamma$ in the RHS of Eq. (\ref{12}) along with the presence of $\Omega$ may seem surprising, since we are dealing with the case $\gamma  \gg \Omega $.  However the longitudinal component $w\sim 1$ has no time to change its value during the pulse owing to relaxation, since $\gamma \tau  \ll 1$.  On the other side, before the pulse the transverse component $v = 0$, and during the pulse it increases its value up to the $v\sim \Omega w\tau $.  Thus the relaxation term in the LHS of Eq. (\ref{8}) remains much smaller than the RHS, $\gamma v \ll \Omega w$, due to the same inequality $\gamma \tau  \ll 1$.

Since the arrival time $T$ considerably exceeds the pulse duration $\tau$, we can shift the lower limit in the integral in Eq. (\ref{12}) to $ - \infty $.  Thence we find that during a short interval around the arrival time the incoming pulse changes the rate of the transition of atom from the excited to the ground state, and this rapid change results in an induced shift of the probability of the excited state by $\Delta P_ +   = {{\Delta w} \mathord{\left/ {\vphantom {{\Delta w} 2}} \right. \kern-\nulldelimiterspace} 2}$, or
\begin{equation}\label{13}
\Delta P_ +   =  - \frac{\pi }{2}\left( {\frac{{d{\mathcal{E}}_0 l}}{{\hbar c}}} \right)^2 w\left( T \right).
\end{equation}
This quantity can serve as a measure of influence of the irradiation by one photon on the process of spontaneous emission.

By using Eqs. (\ref{3}) and (\ref{11}), Eq. (\ref{13}) could be cast into the form 
\begin{equation}\label{14}
\Delta P_ +   =  - \sqrt {\frac{\pi }{8}} \left( {\frac{{\sigma _ 0  }}{S}} \right)\left( {\frac{\gamma }{\delta }} \right)w\left( T \right),
\end{equation}
where 
\begin{equation}\label{15}
\sigma _ 0   = \frac{3}{{2\pi }}\lambda _0^2 
\end{equation}
is the maximal cross-section of the resonant fluorescence of the two-level atom ($\lambda _0  = {{2\pi c} \mathord{\left/ {\vphantom {{2\pi c} {\omega _0 }}} \right. \kern-\nulldelimiterspace} {\omega _0 }}$ is the wave length of the resonant radiation).

Eq. (\ref{14}) shows that the change in the populations of a two-level atom under the influence of a pulse of the classical electromagnetic field that is equivalent to a single-photon wave packet is proportional to the fraction of the transverse section of the pulse that is covered by the cross-section of the resonance fluorescence, and to the fraction of the spectral density of the incoming radiation that gets into the band of the resonant interaction – that is, the natural line width.  This result intuitively seems obvious.

With the parameters chosen above we have $\sigma _ +   = 1.35 \cdot 10^{ - 9} \,\,{\rm{cm}}^2 $, $S = 5 \cdot 10^{ - 3} \,\,{\rm{cm}}^2 $, $\gamma  = 6.70 \cdot 10^6 \,\,{\rm{s}}^{ - 1} $, $\delta  = 10^9 \,\,{\rm{s}}^{ - 1} $, and $\max \left| {\Delta P_ +  } \right| = 2.15 \cdot 10^{ - 9} $, a very small quantity.

\section{ Quantum approach }

Now we turn to study the interaction of atom with the one-photon packet of the quantized electromagnetic field.  For the atom we shall use the same two-level model with the excited state $\left|  +  \right\rangle $ and the ground state $\left|  -  \right\rangle $, which are connected by the electrical dipole transition with the matrix element of the dipole moment ${\bf{d}}$, and the atomic transition frequency $\omega _0 $.  The electromagnetic field is described by the modes of the quantization cube, an imaginary cube with the edge length  with conditions of periodicity imposed on the field on its faces \cite{L73}.  The mode $\mu$ is characterized by its wave vector ${\bf{k}}_\mu  $ and its polarization vector ${\bf{e}}_\mu  $ that obey the condition ${\bf{k}}_\mu  {\bf{e}}_\mu   = 0$, and the mode frequency is $\omega _\mu   = c\left| {{\bf{k}}_\mu  } \right|$.

The Hamiltonian of the system we take in the form
\begin{equation}\label{16}
\hat H = \frac{{\hbar \omega _0 }}{2}\hat \sigma _z  + \sum\limits_\mu  {\hbar \omega _\mu  \hat a_\mu ^ +  \hat a_\mu  }  + i\hbar \sum\limits_\mu  {g_\mu  \left( {\hat \sigma _ +  \hat a_\mu   - \hat a_\mu ^ +  \hat \sigma _ -  } \right)},
\end{equation}
where $\hat \sigma _i $ are the Pauli matrices, $\hat \sigma _ \pm   = {{\left( {\hat \sigma _x  \pm i\hat \sigma _y } \right)} \mathord{\left/ {\vphantom {{\left( {\hat \sigma _x  \pm i\hat \sigma _y } \right)} 2}} \right. \kern-\nulldelimiterspace} 2}$ and $\hat a_\mu ^ +  $
and $\hat a_\mu  $ are the operators of creation and annihilation of photons in the mode $\mu$.  Three terms in Eq. (\ref{16}) represent the Hamiltonian of the atom $\hat H_a $, the Hamiltonian of the field $\hat H_f $ and the interaction term $\hat V$ correspondingly.  The interaction parameter 
\begin{equation}\label{17}
g_\mu   = \sqrt {\frac{{2\pi \omega _\mu  }}{{\hbar {\mathcal{V}}}}} {\bf{de}}_\mu , 
\end{equation}
where $ {\mathcal{V}} = L^3$ is the volume of the quantization cube.  The structure of the interaction term assumes the rotating wave approximation (RWA).

The state vector of the system can be expanded as 
\begin{equation}\label{18}
\left| {\Psi \left( t \right)} \right\rangle  = \sum\limits_\mu  {a_\mu  \left|  +  \right\rangle \left| {1_\mu  } \right\rangle }  + \sum\limits_\mu  {b_\mu  \left|  -  \right\rangle \left| {2_\mu  } \right\rangle }  + \frac{1}{2}\sum\limits_{\mu ,\nu } {c_{\mu \nu } } \left|  -  \right\rangle \left| {1_\mu  } \right\rangle \left| {1_\nu  } \right\rangle ,
\end{equation}
where $\left| {N_\mu  } \right\rangle $ is the $N$-photon Fock state of the mode $\mu$.  The indices in the last sum must not take the equal values.  The vector $\left| \Psi  \right\rangle $ is normalized by the condition
\begin{equation}\label{19}
\left\| \Psi  \right\|^2  = \sum\limits_\mu  {\left| {a_\mu  } \right|^2 }  + \sum\limits_\mu  {\left| {b_\mu  } \right|^2 }  + \frac{1}{2}\sum\limits_{\mu ,\nu } {\left| {c_{\mu \nu } } \right|^2 }  = 1.
\end{equation}

By substituting the expansion Eq. (\ref{18}) in the Schrodinger equation, we obtain the system of equations for the probability amplitudes 
\begin{equation}\label{20}
\frac{{da_\mu  }}{{dt}} =  - \sqrt 2 g_\mu  b_\mu  e^{i\Delta _\mu  t}  - \sum\limits_\nu  {g_\nu  c_{\mu \nu } e^{i\Delta _\nu  t} },
\end{equation}
\begin{equation}\label{21}
\frac{{db_\mu  }}{{dt}} = \sqrt 2 g_\mu  a_\mu  e^{ - i\Delta _\mu  t},
\end{equation}
\begin{equation}\label{22}
\frac{{dc_{\mu \nu } }}{{dt}} = g_\mu  a_\nu  e^{ - i\Delta _\mu  t}  + g_\nu  a_\mu  e^{ - i\Delta _\nu  t}.
\end{equation}
Here the frequency detuning is the difference between the transition frequency and the mode frequency, $\Delta _\alpha   = \omega _0  - \omega _\alpha  $.

We take the initial state of the field in the form of the normalized single-photon wave packet 
\begin{equation}\label{23}
\left| {\Phi \left( 0 \right)} \right\rangle  = \sum\limits_\mu  {\phi _\mu  \left| {1_\mu  } \right\rangle }.
\end{equation}

Then the initial conditions for the system Eqs. (20-22) are
\begin{equation}\label{24}
a_\mu  \left( 0 \right) = \phi _\mu  ,\,\,\,\,\,\,b_\mu  \left( 0 \right) = 0,\,\,\,\,\,\,\,c_{\mu \nu } \left( 0 \right) = 0.
\end{equation}

From the results of the previous section we can conclude that for the typical values of parameters the influence of the incoming photon on the process of the spontaneous emission will be very small and the process will mainly evolve in the same way as in the unperturbed atom.  Then, following the approach of Weisskopf and Wigner \cite{WW30}, we can assume that all amplitudes of the one-photon states decrease with time by the same exponential law with the rate $\gamma  = {\Gamma  \mathord{\left/ {\vphantom {\Gamma  2}} \right. \kern-\nulldelimiterspace} 2}$ that is one half of the rate of the spontaneous decay given by Eq. (\ref{11}):
\begin{equation}\label{25}
a_\mu  \left( t \right) = \phi _\mu  \exp \left( { - \gamma t} \right).
\end{equation}
By substitution of the ansatz Eq. (\ref{25}) in Eq. (\ref{21}) and the integration with the initial conditions Eq. (\ref{24}) we have
\begin{equation}\label{26}
b_\mu  \left( t \right) = \sqrt 2 g_\mu  \phi _\mu  \frac{{1 - e^{ - \gamma t - i\Delta _\mu  t} }}{{\gamma  + i\Delta _\mu  }}.
\end{equation}
Analogously, from Eq. (\ref{22}) we have
\begin{equation}\label{27}
c_{\mu \nu } \left( t \right) = g_\mu  \phi _\nu  \frac{{1 - e^{ - \gamma t - i\Delta _\mu  t} }}{{\gamma  + i\Delta _\mu  }} + g_\nu  \phi _\mu  \frac{{1 - e^{ - \gamma t - i\Delta _\nu  t} }}{{\gamma  + i\Delta _\nu  }}.
\end{equation}

The values of the initial amplitudes depend on the quantization volume.  This dependence can be seen from the normalization condition Eq. (\ref{19}).  For simplification we temporarily ignore the polarization and will consider that the value of ${\bf{k}}_\mu  $ defines completely the mode $\mu$.  Let the initial state of the packet be given by the probability amplitude in the $k$-space $\psi \left( {\bf{k}} \right)$ that is normalized by the condition
\begin{equation}\label{28}
\int {\left| {\psi \left( {\bf{k}} \right)} \right|^2 } d{\bf{k}} = 1.
\end{equation}
For large enough volume $\mathcal{V}$ the summation over $\mu$ could be replaced by the integration over $\bf{k}$
\begin{equation}\label{29}
\sum\limits_\mu  {\left| {\phi _\mu  } \right|^2 }  = \sum\limits_\mu  {\left| {\phi \left( {{\bf{k}}_\mu  } \right)} \right|^2 }  = \int {\left| {\phi \left( {\bf{k}} \right)} \right|^2 \rho \left( {\bf{k}} \right)} d{\bf{k}}.
\end{equation}
where $\rho \left( {\bf{k}} \right)$ is the density of permissible discrete values of the vector ${\bf{k}}_\mu  $ in the $k$-space,
\begin{equation}\label{30}
\rho \left( {\bf{k}} \right) = \frac{\mathcal{V}}{{\left( {2\pi } \right)^3 }}.
\end{equation}
Thus from comparison of Eqs. (\ref{28}) and (\ref{29}) the discrete amplitudes will have the form 
\begin{equation}\label{31}
\phi _\mu   = \sqrt {\frac{{8\pi ^3 }}{\mathcal{V}}} \psi \left( {{\bf{k}}_\mu  } \right).
\end{equation}

The summary probability of the doubly occupied field modes,
\begin{equation}\label{32}
P_2  = \sum\limits_\mu  {\left| {b_\mu  } \right|^2 }  \propto g^2 \phi ^2 \rho  \propto {\mathcal{V}}^{ - 1},
\end{equation}
in the limit ${\mathcal{V}} \to \infty $ turns to zero.  Thus the probability of creation of doubly occupied state in the process of interaction of a single-photon wave packet with the excited atom in the free space is zero.  This result is quite general: it is not limited by the applicability of the ansatz Eq. (\ref{25}), since one can substitute Eqs. (\ref{26}) and (\ref{27}) into Eq. (\ref{20}), find a new approximation for $a_\mu  $ and continue the iterations, and on every step the scaling relation Eq. (\ref{32}) will hold.

The summary probability of pairs of singly occupied states,
\begin{equation}\label{33}
P_{11}  = \sum\limits_{\mu ,\nu } {\left| {c_{\mu ,\nu } } \right|^2 }  \propto g^2 \phi ^2 \rho ^2  \propto {\mathcal{V}}^0 ,
\end{equation}
does not depend on the volume of quantization and thus gives a finite value for the process in a free space.

\section{ Numerical example }

In principle the expressions Eqs. (\ref{26}) and (\ref{27}) along with the conversion rule Eq. (\ref{31}) present the answer to the problem of evolution of the system in quadratures.  However the numerical integration over two copies of the $k$-space in three-dimensional case happens to be too demanding on the computer resources.  To demonstrate the properties of the solution we limit ourselves by a one-dimensional case that still have some direct physical interest.

Let's assume that the field is located on the interval from ${{ - L} \mathord{\left/ {\vphantom {{ - L} 2}} \right. \kern-\nulldelimiterspace} 2}$ to ${L \mathord{\left/ {\vphantom {L 2}} \right. \kern-\nulldelimiterspace} 2}$ with periodic boundary conditions.  In the following we will use the system of units in which the average wave number of the incoming packet $K = 1$; we also take $\hbar  = 1$ and $c=1$.  We shall replace the interaction parameters $g_{\mu}$ by the constant value $g$; that is relatively unimportant since only components around the transition frequency take active part in the interaction.  We also limit ourselves to a single polarization.  The connection between the interaction constant $g$ and the decay rate $\gamma$ follows from the Fermi golden rule: $\gamma  = g^2 L$.

We take the packet of the Gaussian form
\begin{equation}\label{34}
\phi _\mu   = \phi \left( {k_\mu  } \right) = \frac{{\left( {2\pi } \right)^{{1 \mathord{\left/ {\vphantom {1 4}} \right. \kern-\nulldelimiterspace} 4}} }}{{\sqrt {\kappa L} }}\exp \left[ { - \frac{{\left( {k_\mu   - 1} \right)^2 }}{{4\kappa ^2 }} - ik_\mu  \Lambda } \right].
\end{equation}

From Eqs. (\ref{27}) and (\ref{34}), using the one-dimensional analogs of Eqs. (\ref{30}) and (\ref{31}), we obtain the expression for the time-dependent probability density of photons in the $k$-space:
\begin{equation}\label{35}
C\left( {k_1 ,k_2 ;t} \right) = \frac{1}{{\sqrt {8\pi ^3 } }}\frac{\gamma }{\kappa }\left| {\xi \left( {k_1 } \right)\chi \left( {k_2 ,t} \right) + \xi \left( {k_2 } \right)\chi \left( {k_1 ,t} \right)} \right|^2 ,
\end{equation}
where
\begin{equation}\label{36}
\xi \left( z \right) = \exp \left[ { - \frac{{\left( {z - 1} \right)^2 }}{{4\kappa ^2 }} - iz\Lambda } \right],
\end{equation}
\begin{equation}\label{37}
\chi \left( {z,t} \right) = \frac{{1 - \exp \left[ { - \gamma t - i\left( {1 - \left| z \right|} \right)t} \right]}}{{\gamma  + i\left( {1 - \left| z \right|} \right)}}.
\end{equation}
The total probability of finding the atom in the ground state and two photons in the field is given by the twofold integral,
\begin{equation}\label{38}
P_ -  \left( t \right) = { \frac {1} {2}} {\int \int {C\left( {k_1 ,k_2 ;t} \right)dk_1 dk_2 }},
\end{equation}
where the integration is carried out over all values of $k_1$ and $k_2$.

For the numerical calculations we choose the values of parameters $\gamma = 0.0125$ and $\kappa = 0.25$.  This value of the wavenumber distribution width corresponds to the single-photon packet with the spatial length $l = \kappa ^{ - 1}  = 4$ and the spectral width $\delta  = c\kappa  = 20\gamma $.  The arrival time $T =  - \Lambda $ will take different values; we will indicate them in $\Delta P_ +  \left( T \right)$.

The numerically found dependence of $P_ -  \left( t \right)$ is shown in Fig. 1.   For the negative displacement $\Lambda  =  - 20$ around the arrival tine $T = 20$ the growth of the probability $P_-$ accelerates for a short time in accordance to the semiclassical description given in Sec. 2.  For comparison the results for the positive displacement $\Lambda  =  20$ are shown.  In the latter case the packet has passed the atom before it become excited and never hit it.  Naturally, the dependence shows the behavior appropriate to the spontaneous emission of the unperturbed atom.  The vertical distance between the curves in the right part of the graph in Fig. 1 represents the induced probability shift; its calculated value is $\Delta P\left( {20} \right) =  - 0.076$.

\begin{figure}[!ht]
\includegraphics[width=0.6\columnwidth]{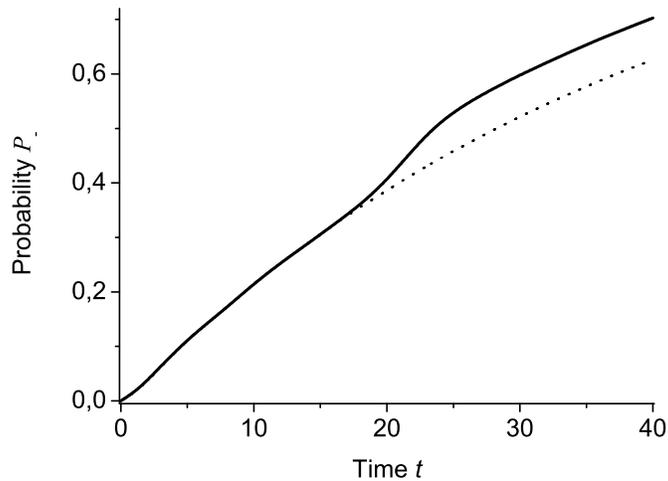}
\caption{\label{fig1} The dependence of the probability $P_-$ of finding the atom in the ground state on time $t$ for values of the initial displacement $\Lambda  =  - 20$ (solid line), and $\Lambda  = 20$ (dash-dotted line).}
\end{figure}

The time-dependent rate of the downward transition can be defined as
\begin{equation}\label{39}
\Gamma \left( t \right) = - \frac{1}{{P_ +  }}\frac{{dP_ +  }}{{dt}},
\end{equation}
where $P_ +  \left( t \right)$ is the probability of finding atom in the excited state.  The behavior of $\Gamma \left( t \right)$ is shown in Fig. 2.  The maximum of the decay rate is reached at $t = 21.6$, that is somewhat later than the arrival time $T = 20$.  The maximal value of $\Gamma$ exceeds the rate of transition due to the spontaneous emission by the factor 1.94.

\begin{figure}[!ht]
\includegraphics[width=0.6\columnwidth]{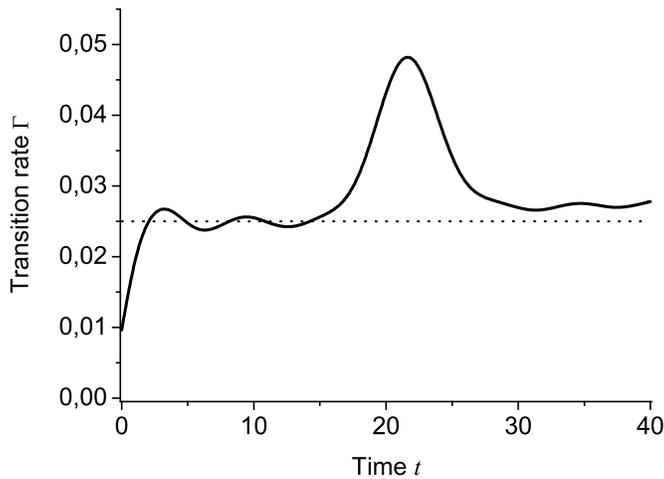}
\caption{\label{fig2} The dependence of the temporal rate of the downward transition $\Gamma$, defined by Eq. (\ref{39}), on time $t$ for the initial displacement $\Lambda  =  - 20$ (solid line).  The dashed line shows value of $\Gamma  = 2\gamma  = 0.025$ for the unperturbed atom.}
\end{figure}

If we apply the semiclassical approximation presented in Sec. 1, then for the one-dimensional case (cf. Eq. (\ref{12})) we obtain for the induced shift of the probability
\begin{equation}\label{40}
\Delta P_ +   =  - \sqrt {2\pi } \frac{\gamma }{\delta }w\left( T \right).
\end{equation}
In our numerical example this formula yields $\Delta P\left( {20} \right) =  - 0.027$, that is about one third of the value given by the quantum calculation.  What is more important, Eq. (\ref{40}) predicts the change of sign of the induced probability shift for $T \ge \left( {{{\ln 2} \mathord{\left/ {\vphantom {{\ln 2} 2}} \right. \kern-\nulldelimiterspace} 2}} \right)\gamma ^{ - 1} $, that is, slowing down of the process of spontaneous radiation.  The quantum calculation shows that the incoming photon always accelerate the downward transition irrespectively of the sign of the population difference $w\left( T \right)$; instead of Eq. (\ref{40}) it yields
\begin{equation}\label{41}
\Delta P_ +   =  - \sqrt {2\pi } \frac{\gamma }{\delta }P_ +  \left( T \right).
\end{equation}
The reason of the proportionality of the quantum value of $\Delta P_ +  $ to the population of the upper state at the time of the arrival $P_ +  \left( T \right)$ is clear: with the ansatz Eq. (\ref{25}) the behavior of the amplitudes $a_\mu  \left( t \right)$ does not depend  on the parameters of the wave packet at all.  The choice of positive $T$ just uniformly decreases values of all amplitudes by multiplying them by the factor $\exp \left( { - \gamma T} \right)$.

At this point we have to admit the superiority of the semiclassical theory over the simplest approximate solution of the quantum model.  Let's consider the case $2\gamma T \gg 1$, in which the packet arrives to the atom, when it is almost exactly in the ground state.  In this case the action of the photon on the atom will not depend on $T$.  The incoming photon will excite the atom with some probability $\Delta P_ +  \left( \infty  \right)$ (just because there is no other way of evolution from the state with $w =  - 1$), then the process of spontaneous emission will be renewed from the initial value $\Delta P_ +  \left( \infty  \right)$.  This qualitative picture is in agreement with the semiclassical result Eq. (\ref{40}).

The quantity $\Delta P_ +  \left( \infty  \right)$ can be calculated from a purely quantum model.  Taking the expansion of the state vector in the form
\begin{equation}\label{42}
\left| {\Psi \left( t \right)} \right\rangle  = A\left|  +  \right\rangle \left| {{\rm{vac}}} \right\rangle  + \sum\limits_\mu  {B_\mu  \left|  -  \right\rangle \left| {1_\mu  } \right\rangle } 
\end{equation}
and its substitution in the Schrodinger equation with the Hamiltonian Eq. (\ref{16}) lead to the system of equations for the probability amplitudes
\begin{equation}\label{43}
\frac{{dA}}{{dt}} =  - g\sum\limits_\mu  {B_\mu  e^{i\Delta _\mu  t} } ,\,\,\,\,\,\,\,\,\,\,\frac{{dB_\mu  }}{{dt}} = gAe^{ - i\Delta _\mu  t}.
\end{equation}
The numerical solution of this system with the chosen values of $\kappa $ and $\gamma $ has been carried out for the length of the interval of quantization $L = 251.32$ with the account of $\mathcal{N} = 159$ modes with minimal $\left| {k_\mu  } \right|$.  When tested, this approximation has produced the value of the spontaneous decay rate that differed from the theoretical value by less than 1\%.  The calculated dependence of the $P_ +  \left( t \right) = \left| {A\left( t \right)} \right|^2 $ is shown in Fig. 3.

\begin{figure}[!ht]
\includegraphics[width=0.6\columnwidth]{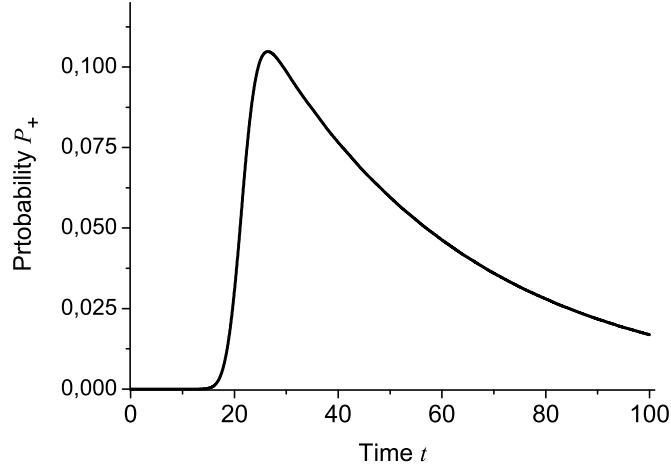}
\caption{\label{fig3} The dependence of probability of the excited state $P_+$ on time $t$ for the case of the one-photon packet Eq. (\ref{34}) with the initial displacement $\Lambda  =  - 20$ scattered on the atom in the ground state.}
\end{figure}

The numerically found quantum value $\Delta P_ +  \left( \infty  \right) = 0.105$ is quite close to the semiclassical value $\Delta P_ +  \left( \infty  \right) = 0.125$ that follows from Eq. (\ref{40}).

The constancy of the negative sign of $\Delta P_ +  \left( T \right)$ that follows from the simplest approximate quantum solution can be improved by the iteration process.  By substitution of the solution Eq. (\ref{27}) in Eq. (\ref{20}) (the first term can be neglected in accordance with Eq. (\ref{32})) we obtain the formula for the amplitudes in the approximation of the first order,
\begin{equation}\label{44}
a\left( {k_\mu  ,t} \right) = \phi \left( {k_\mu  } \right) - \frac{\gamma }{{2\pi }}\int\limits_{ - \infty }^{k_\mu  } {F\left( {k_\mu  ,k_\nu  ;t} \right)\,} dk_\nu  ,
\end{equation}
where $\phi \left( z \right)$ is given by Eq. (\ref{34}), and

\begin{eqnarray}\label{45}
F\left( {x,y;t} \right) = \frac{{\phi \left( x \right)}}{{\gamma  + i\left( {1 - \left| x \right|} \right)}}\left\{ {G\left[ {i\left( {1 - \left| y \right|} \right)} \right] - G\left[ { - \gamma  + i\left( {\left| x \right| - \left| y \right|} \right)} \right]} \right\} +  \\ 
\nonumber \,\,\,\,\,\,\,\, + \frac{{\phi \left( y \right)}}{{\gamma  + i\left( {1 - \left| y \right|} \right)}}\left\{ {G\left[ {i\left( {1 - \left| y \right|} \right)} \right] - G\left[ { - \gamma } \right]} \right\}, 
\end{eqnarray}
\begin{equation}\label{46}
G\left( z \right) = \frac{{e^z  - 1}}{z}.
\end{equation}
These expressions permit to calculate directly the kinetics of the probability of the excited state 
\begin{equation}\label{47}
P_ +  \left( t \right) = \frac{L}{{2\pi }}\int\limits_{ - \infty }^\infty  {\left| {a\left( {k,t} \right)} \right|^2 dk}.
\end{equation}
In the first approximation $\Delta P_ +  \left( {20} \right) =  - 0.059$, that is closer to the semiclassical value than the result of the zeroth approximation.  The value of the $\Delta P_ +  $ calculated from the expression Eq. (\ref{47}) for large $T$ becomes positive (e.g. $\Delta P_ +  \left( {130} \right) = 3 \cdot 10^{ - 3} $), although it remains much smaller than the numerically found limiting value $\Delta P_ +  \left( \infty  \right) = 0.105$.  One may hope that the higher approximations will improve the agreement.

Now we discuss the spectral properties of the photons in the final state.  The spectral density of the final photons $S\left( \omega  \right)$ can be expressed through the limiting value of the double $k$-space density given by Eq. (\ref{35}), $C_f \left( {x,y} \right) = C\left( {x,y;\infty } \right)$.  The expression has the form
\begin{equation}\label{48}
S\left( \omega  \right) = \int\limits_0^\infty  {\,\left[ {C_f \left( {\omega ,\omega '} \right) + C_f \left( { - \omega ,\omega '} \right) + C_f \left( {\omega , - \omega '} \right) + C_f \left( { - \omega , - \omega '} \right)} \right]d\omega '}.
\end{equation}
Since the interaction of the photon with the atom is weak, it is natural to assume that the spectrum is close to the sum of two relevant spectra, for $\delta  \gg \gamma $ - the wide Gaussian form of the incoming packet and the narrow Lorentzian form of the spontaneous radiation of the isolated atom,
\begin{equation}\label{49}
S_0 \left( \omega  \right) = \frac{1}{{\sqrt {2\pi } \delta }}\exp \left[ { - \frac{{\left( {\omega  - 1} \right)^2 }}{{2\delta ^2 }}} \right] + \frac{1}{\pi } \cdot \frac{\gamma }{{\gamma ^2  + \left( {1 - \omega } \right)^2 }}.
\end{equation}
We shall call this expression the basic spectrum.  The numerical calculations show that the final spectrum of photons is very close to this form indeed (see Fig. 4).

\begin{figure}[!ht]
\includegraphics[width=0.6\columnwidth]{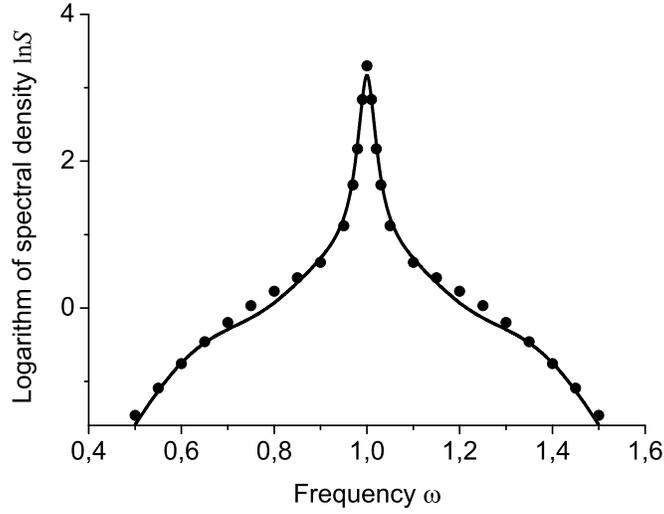}
\caption{\label{fig4} The logarithm of the spectral density as a function of frequency $\omega $.  The solid line shows the calculated numerically values of $S$ for the initial displacement $\Lambda  =  - 20$.  The dots shows the basic spectrum $S_0$.}
\end{figure}

It is difficult from this picture to see the deviations of the real spectrum from the summary one that are created by the interaction of the single-photon wave packet with the excited atom.  The ratio of spectral densities that is shown in Fig. 5 is more instructive.

\begin{figure}[!ht]
\includegraphics[width=0.6\columnwidth]{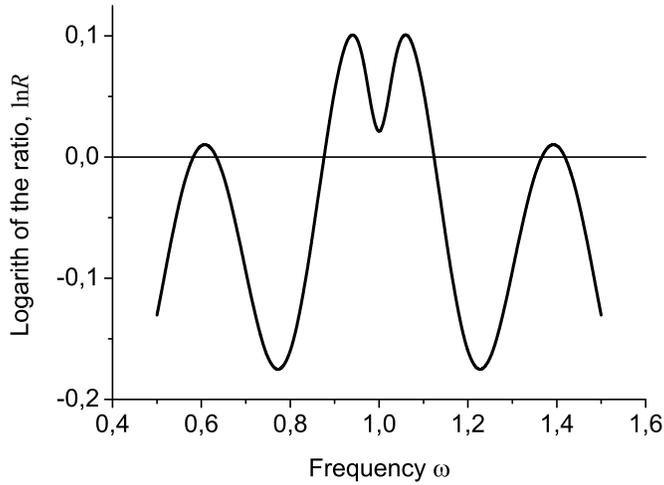}
\caption{\label{fig5} The logarithm of the ratio of the real spectral density (found numerically) to the summary one, $R =  {S}/ {S_0}$, as a function of frequency $\omega $.}
\end{figure}

The interaction somewhere increases the spectral density and somewhere decreases it.  We define the spectral with $\Delta \omega $ as a minimal interval that contains half of the total spectral density.  With this definition we find that the net effect of the interaction produces small, 11\%, broadening of the spectrum.

\section{Discussion}

Einstein in his groundbreaking papers of 1916/17 \cite{E16, E16a, E17} treated both radiation induced processes - that of absorption and that of acceleration of the downward transition - on equal footing.  He named them ''Zustands\"{a}nderungen durch Einstrahlung'', changes of state due to irradiation.  The specific term for the downward process, ''negative Einstrahlung'', negative irradiation, came in use by 1923 \cite{EE23}.  In 1924 Tolman \cite{T24} introduced the term ''negative absorption'', and in the same year van Vleck launched the parallel term ''induced emission'' \cite{vV24}.  Finally, the predominant at the present time term ''stimulated emission'' was brought in by Dirac \cite{D27}.  This chain of renamings has shifted the focus of attention from the evolution of the atomic (molecular) system to that of the radiation.  Nevertheless, it is the acceleration of the downward transition in the atomic system induced by irradiation that remains the determinative indicator of the stimulated emission.

Dirac established the theory of interaction of atoms with the quantized electromagnetic field and derived Einstein's phenomenological equations from this theory \cite{D27}.  In the course of derivation it became clear, that the specific property of the operator of creation of photons,
\begin{equation}\label{50}
\hat a^ +  \left| N \right\rangle  = \sqrt {N + 1} \left| {N + 1} \right\rangle, 
\end{equation}
is responsible for the acceleration.  Dirac used the approximation that is known now as the Fermi golden rule: the rate of the atomic transition is proportional to the square of modulus of the matrix elements summed over different modes.  On the other hand, Eq. (\ref{50}) can be interpreted as a statement that the stimulated photons are radiated to the same mode (definitely with the same frequency $\omega _\mu  $, and if the modes are taken in the form of travelling plane waves - then with the same wave vector ${\bf{k}}_\mu  $ and the polarization ${\bf{e}}_\mu  $), making the exact copies of the initial, stimulating photon(s).

In Dirac's approximation the acceleration of the transition, that is the stimulated emission, is due to the process of emitting photons into the occupied modes, that can be interpreted as a process of radiation of exact copies of the initial photons.  In the course of history the fine print was frequently neglected, and the previous (correct) statement has been sometimes reduced to assertion that ''stimulated emission is a process of radiation of exact copies of the initial photons.''  This statement may be still correct, if we make it a definition and will apply the term ''stimulated emission'' only to vectors of multiply occupied modes.  Some authors do that \cite {WZ82, SL+07}.  But one must remember that this definition is different from the historically developed one.

Acceleration of the transition can occur without copying of photons.  Drobny, Havukainen, and Buzek \cite{DHB00} studied the scattering of a single-photon wave packet on an atom in the \textit {ground} state.  In this situation all radiation is emitted into vacant modes, since in the RWA, that was used by the authors, with the given initial conditions the creation of the doubly occupied mode is strictly forbidden.  However, the temporal rate of the transition from the upper state (see Eq. (\ref{39})) in their studies reached values about an order of magnitude larger than that of the spontaneous emission of the unperturbed excited atom.  In the problem studied in the present paper, that of scattering of a single-photon wave packet on an atom in the \textit{excited} state, the situation is similar: the acceleration may be present (see Fig. 2), whereas the copying is absent (see Eq. (\ref{32})), albeit not due to the limitations of the RWA.

The results presented above could be verified experimentally.  A single atom in a trap could be excited by a short pi-pulse right before the arrival of a single-photon wave packet.  The most favorable conditions for spectral studies correspond to the case in which the spectral width of the packet $\delta $ is somewhat larger than the natural line width $\gamma $, but not too much (for instance, $\delta  = 20\gamma $ as in the numerical example of Sec. 4).  Then the accuracy of measurement of the spectral density about 1\% will be enough to observe the deviations from the summary spectrum shown in Fig. 4.  Furthermore, at present the methods of experimental study of multiple occupancy of modes are known \cite{H09}, and the absence of the multiply occupied modes could be confirmed directly.

The problem solved in this paper may be of some interest for studies of the cloning of photons \cite{WZ82,SI+05}, especially of theirs non-polarizational degrees of freedom.

\section{acknowledgement}
The author is grateful to G.A. Chizhov, S.P. Kulik and E.A. Ostrovskaya for useful discussions, informational support and wise advice.  The author acknowledges the support by the "Russian Scientific Schools" program (grant NSh 4464.2006.2) and by the Russian Foundation for Basic Research (grant \#11-02-00317-A).

\end{document}